# III-V-based optoelectronics with low-cost dynamic hydride vapor phase epitaxy

John Simon, Kelsey Horowitz, Kevin L. Schulte, Timothy Remo, David L. Young, and Aaron J. Ptak

National Renewable Energy Laboratory, Golden, CO 80401, USA

**Silicon is the dominant semiconductor in many semiconductor device applications for a variety of reasons, including both performance and cost. III-V materials have improved performance compared to silicon, but currently they are relegated to applications in high-value or niche markets due to the absence of a low-cost, high-quality production technique. Here we present an advance in III-V materials synthesis using hydride vapor phase epitaxy that has the potential to lower III-V semiconductor deposition costs while maintaining the requisite optoelectronic material quality that enables III-V-based technologies to outperform Si. We demonstrate the impacts of this advance by addressing the use of III-Vs in terrestrial photovoltaics, a highly cost-constrained market.**

Silicon as a semiconductor technology is beginning to run into significant technical limits. The death of Moore's Law has been predicted for decades, but there is now clear evidence that transistor size limits have been reached, and improvements are only being realized through increases in complexity and cost. Si is also rapidly approaching the practical limit for solar conversion efficiency with current best performance sitting at 26.6%[1]. In contrast to Si, III-V materials such as GaAs and GaInP have some of the best electronic and optical properties of any semiconductor materials. III-Vs have higher electron mobilities than Si, enabling transistors operating at high frequencies for wireless communication applications, and direct bandgaps that lead to extremely efficient absorption and emission of light. These materials appear, among other places, in power amplifiers that enable transmitting and receiving capabilities in cell phones, as high-value space-based photovoltaic (PV) panels, and in light emitting diodes (LEDs) for general illumination applications with nitride-based III-V. III-V PV devices hold record conversion efficiencies for both single [2] and multi-junction [1] solar cell devices, as well as one-sun modules[1]. Unlike Si, they can be quite thin and flexible while maintaining high conversion efficiency; they can reject heat, permitting them to operate at lower temperatures in real world outdoor conditions [3]; and they have lower temperature coefficients [4], resulting in minimal performance degradation when their temperature does rise, which can reduce the requirements on heat sinking [3] and allow solar cells to be in intimate contact with rooftops. III-V materials are readily integrated in multijunction solar cell structures that increase efficiency far beyond single junction limits. These qualities allow III-V PV modules to produce more energy than a similarly power rated silicon PV module over the course of their lifetime.[3]

The development of III-V materials and devices historically focused on quality, efficiency, and performance, with less regard to the cost of the epitaxial growth, and III-Vs lacked a driving force like Si CMOS to methodically push manufacturing significantly costs lower. So, while the performance of III-V devices is undeniably excellent, their cost has limited their use to applications where the characteristics of the III-V materials are necessary to achieve required performance, and/or the high cost of the manufacturing is amortized over the many devices grown in a single batch deposition run. For example, thousands of LEDs, laser diodes, monolithic microwave integrated circuits, and heterojunction bipolar transistors are produced during one growth run in a production III-V reactor. In large-area applications like PV, where costs cannot be spread over numerous devices per batch, III-Vs are currently only used in niche, high-value (and low volume) markets such as space power, concentrating PV in areas with high direct normal irradiance, and more recently, in area- and weight- constrained applications like unmanned aerial vehicles (UAVs). If III-V materials were produced more cheaply than is possible using today's manufacturing techniques, more widespread adoption of III-V's in PV and other opto-electronic applications could be achieved, and this increased market presence can further reduce manufacturing costs similar to what was seen in Si cost reductions as it expanded into various applications.

Thus, an innovative III-V manufacturing process is required. Hydride vapor phase epitaxy (HVPE) is a semiconductor growth technique that combines high epitaxial growth quality, high throughput, and high precursor material utilization. Several key features make HVPE more cost effective than current III-V epitaxial growth processes, including deposition rates as high as 300 µm/h for GaAs[5], the use of low-cost, elemental metal sources in the reaction, and high utilization of the source materials, particularly hydride gases. HVPE was developed in the 1960s and was used commercially for the production of GaAsP LEDs, as well as photo-emitters and photo-detectors for the telecommunications industry. HVPE largely fell out of favor, however, due to technical challenges not experienced by today's incumbent technologies, such as metal organic vapor phase epitaxy (MOVPE) and molecular beam epitaxy (MBE). Despite the obvious potential cost benefits delivered by HVPE, the high speed of the growth process and the residence times of the process gases made it difficult to achieve the low-defect and chemically-abrupt heterointerfaces critical in many device structures.

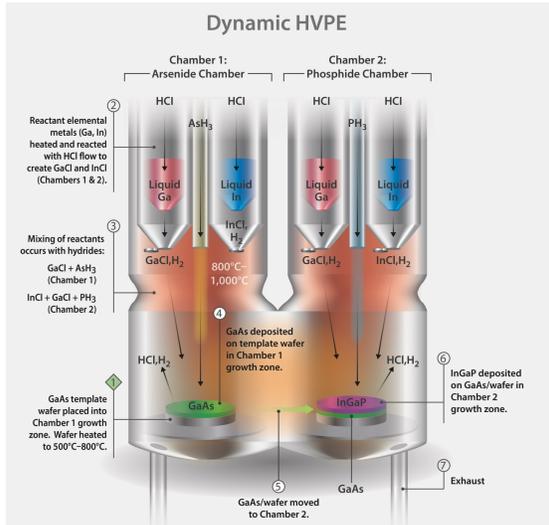

Fig. 1. Schematic of the two-chamber D-HVPE reactor at NREL showing side-by-side steady-state reactions for GaAs and GaInP. The substrate is rapidly shuttled between the growth chambers to create high-quality, chemically abrupt interfaces.

To enable abrupt heterointerfaces vital to high performance III-V devices, but still maintain high throughput, we developed dynamic hydride vapor phase epitaxy (D-HVPE), a new route to low-cost III-V growth. D-HVPE enables the creation of abrupt heterointerfaces while maintaining fast deposition rates. This can only be achieved in traditional HVPE by introducing detrimental growth interrupts. We designed a new HVPE reactor with multiple growth chambers, separated by inert gas curtains, shown schematically in Fig. 1. In this approach, abrupt heterointerfaces are formed through translation of a substrate from one growth chamber to another, each of which has an independently established, steady-state deposition reaction for the III-V material to be grown. This technique is effective at generating atomically- and chemically-abrupt interfaces[6,7]. This is the first step toward effective device formation at these high growth rates, and below we will discuss the optical and electrical properties of these interfaces. This two-chamber design takes advantage of the low-cost source materials (elemental metals) and high growth rates inherent to HVPE while enabling abrupt, high quality interfaces. This research scale reactor uses a single 2" substrate for simplicity, but future production scale reactors, like the ones assumed in the next section, will implement multiple, large-area substrates. The current growth system design approximates an in-line production reactor in that the different device layers are deposited as the substrate is shuttled through the various regions of a linear reactor. This design allows evaluation of the effectiveness of in-line device manufacturing without the need to develop a fully in-line reactor. Employment of an in-line deposition process will provide a pathway towards significant throughput increases and associated cost reductions, similar to how in-line deposition techniques already provide low-cost fabrication of thin-film PV devices, e.g. CdTe. Fig. 2 shows a schematic of what the growth process will look like in a multi-chamber HVPE reactor that is capable of growing a high-efficiency single-junction GaAs solar cell in minutes instead of the hours that it takes with current manufacturing technology.

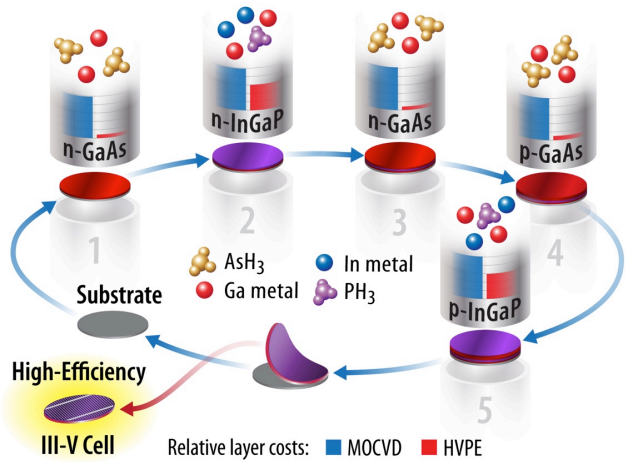

Fig. 2: Schematic of in-line HVPE reactor used to grow low-cost single junction GaAs solar cells together with a substrate reuse technology.

In the next sections, we examine the economic viability of D-HVPE and then demonstrate that this technique can be used to produce devices with comparable performance to MOVPE grown devices. We develop a cost model for the HVPE growth of III-V devices using a linear, in-line deposition system conceptually similar to that shown in Fig. 2. The cost analysis shows that it is possible for III-Vs to be dramatically less expensive than they are today. We focus on the specific case of using D-HVPE for the production of III-V-based solar cells, which can enable competitiveness in various cost-sensitive terrestrial PV applications with market sizes that increase as substrate costs, which are concurrently the subject of intense research, are reduced. In the final section, we describe current efforts to produce defect-free interfaces and high-efficiency single-junction GaAs solar cells using our custom D-HVPE growth system as an example of devices possible using this technique. We demonstrate solar conversion efficiencies > 25.3% in initial single-junction GaAs device grown at ~ 60 µm/h, similar to the efficiency of devices grown using traditional growth technology. We also show initial results for monolithic, HVPE-grown, two-junction GaInP/GaAs solar cell devices that incorporate three separate electrical junctions in a single device. The combination of low cost growth with the demonstration of high efficiency devices illustrates the promise of the D-HVPE technique, not only for solar cells, but for *all* III-V device applications.

**HVPE deposition costs**
We performed an analysis of the potential impact of the HVPE process on III-V deposition costs using NREL's established bottom-up methodology[8-11]. For this analysis, we assume that the NREL two-chamber HVPE reactor is scaled up to a high-volume, continuous reactor with one zone for heating the substrate and one deposition zone per layer, as illustrated schematically in Fig. 1 of the Methods section. The deposition zones are isolated using buffer sections incorporating inert gas curtains. Because no such high volume HVPE reactor exists today, we created a basic model of the system to estimate throughput and cost per tool; this model has been reviewed by members of industry and their feedback incorporated for accuracy. Details of the model appear in the Methods section and in ref 12.

**Cost of HVPE-grown III-V photovoltaic devices**
We focus more specifically on the use of HVPE to produce low-cost, high-efficiency, III-V solar cells and their potential competitiveness in different markets. If commercialized, D-HVPE could immediately provide value to PV markets requiring high efficiency, high specific power, or flexible form factors, including consumer electronics, UAVs, military, space markets, and automotive roofs[13] by reducing the cost of epitaxy while providing similar performance to what is available today. The overall benefit depends on the cell type, production volume, and processes used for other aspects of the cell fabrication (e.g. metallization, choice of substrate) for a given manufacturer.

As discussed above, III-V solar cells at one-sun have not previously penetrated mainstream PV power markets due to their prohibitively high cost. In this section, we assess the potential for D-HVPE and substrate reuse to enable III-V technology to compete in some of these markets by modeling high volume costs for III-V solar cells fabricated using these processes. We explore their potential balance-of-system [BOS] (e.g. racking materials, installation costs) and levelized cost of energy [LCOE] (e.g. cost to produce a kWhr of electricity) advantages over incumbent Si flat plate technology.

Our cell cost models indicate that, at scale, dual junction InGaP/GaAs cells deposited via HVPE could potentially reach costs below $0.50/W, even with U.S. manufacturing. This would allow these solar cells to be competitive in larger PV markets that would benefit from the high power density, low operating temperature and temperature coefficient, and lightweight, flexible form factor provided by III-V materials. Applications could include PV tile roofs, PV on electric vehicles (EVs), and certain residential and commercial rooftop installations that are weight- or area-constrained. In fact at <$0.50/W, III-V solar cells may even be competitive when dropped into traditional PV module and system designs. In the remainder of this section, we explore the case of residential rooftop systems, which have higher BOS costs and areal constraints than typical ground-mount, utility-scale installations, and thus stand to benefit more significantly from the increased efficiency associated with III-V devices. We compare the total installed system cost and LCOE for incumbent monocrystalline Si PERC[14] technology to that of HVPE-deposited III-V cells with substrate reuse. PERC cells were chosen for comparison because they are rapidly gaining traction and are anticipated to become market-dominant over the next few years[15]. We use modeled cell and module prices for both technologies that include overhead costs and a sustainable product margin, rather than using current Si PERC cell prices for the Si case, in order to obtain a technology-based comparison. The details and assumptions of this model appear in the Methods section. The results are shown in Fig. 3. While the III-V single junction cell costs are higher than for PERC cells, the increased efficiency of the III-V single junction cells compared to Si results in balance-of-module [BOM] (e.g. glass, encapsulant, busbars) and BOS cost savings, resulting in comparable total installed system costs. The savings is higher for the dual junction cells due to their higher efficiency, resulting in total installed system costs that could be comparable to those of current PERC technology. At comparable installed system costs, III-V cells should provide a lower LCOE than Si, due to the higher energy yield resulting from their lower operating temperature and temperature coefficient. In prior field measurements, single-junction GaAs cells exhibited 8% higher energy yield than Si cells in Phoenix, AZ in an open-rack configuration, though this will vary with location.[3] Any increases in energy production translate to decreases in LCOE; thus, III-Vs could have an 8% lower LCOE compared to monocrystalline Si in Phoenix. The relative installed cost of PERC cells compared to HVPE-deposited III-Vs is similar for commercial rooftop systems, so similar LCOE reductions would also occur in these markets. Finally, the energy yield improvement and thus LCOE benefit would be even greater in applications like solar shingles where cells are direct-mounted on to the roof because of the lack of a suitable heat sink for the cells.

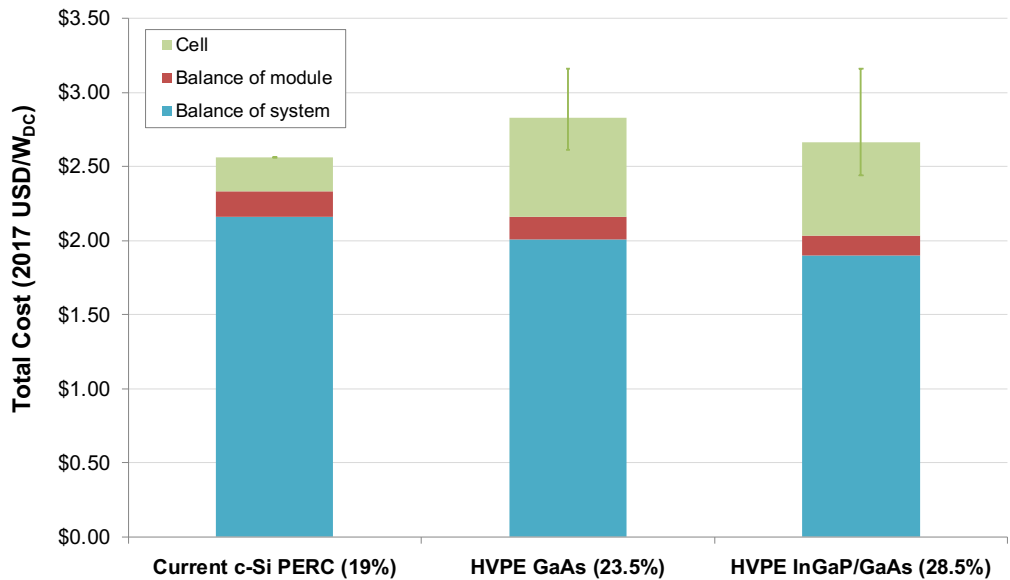

Fig. 3: Comparison of total installed system costs for a 5.6kW residential rooftop system for current monocrystalline Si PERC cells compared to estimates for D-HVPE-deposited GaAs and InGaP/GaAs cells in the case where substrate cost are addressed. Assumes Chinese manufacturing and high-volume production (500 MW/year) for all cases. Cell, module, and system costs shown include overhead costs and sustainable product margins.

The ability to reach costs on the order of $0.50/W depends critically on the cost of the substrate. These cost models assume that the cost of the substrate could be reduced significantly in the long-term to a value of ~ $1.00 per 6" wafer, in this case via a large number (approximately 100, depending on the future GaAs substrate price) of substrate reuses and avoidance of CMP cost. The ability to reuse the substrate with limited or no CMP through the use of a series of buffer layers has been demonstrated in the literature, but this has not yet been demonstrated at scale or with hundreds of wafer reuses,[16,17] and this may be challenging to achieve with high yield. However, active research is taking place on substrate reuse, including chemical lift-off and mechanical fracture technologies, and lower cost substrates. To reach cell costs on the order of $1/W to $5/W, which may be acceptable in some markets (e.g. PV on EVs, other portable power applications), less aggressive substrate cost targets would be sufficient. While there are III-V solar cell companies that currently implement epitaxial lift-off[18] and reuse the substrate, it is unclear how many times the substrate is being reused. Additional research and development is required to demonstrate a hundred reuses at scale with high yield and significantly reduced polishing and reclaim costs. The lift-off process itself must also be scaled-up; current production volumes for III-V cells are low. Alternatively, cost reductions could be achieved via the use of a low-cost substrate (e.g. a virtual substrate or template, direct-growth on a low-cost substrate), as long as similar efficiencies can be obtained, or a combination of the two (lower cost substrate that is reused a lower number of times).

The cost model presented here does assume the immediate implementation of the technology presented in this paper. In addition to the time to commercialize and scale-up the D-HVPE process, the substrate advances will require time and investments in research and development. However, because the results in Fig. 3 are based on 2017 module and system cost structures, and include modeled 2017 Si PERC cell costs, these comparisons essentially assume these III-V cell costs were achieved overnight. While this serves to illustrate the benefits of higher III-V cell efficiency at the module and system level and provides a useful benchmark for understanding whether or not III-V cells might be applicable to general power markets, some additional discussion of the future is warranted. Further BOM and BOS cost reductions are anticipated in the future.[19] These reductions would benefit both Si and III-Vs, but the marginal value of higher efficiencies, and thus the advantage of III-V cells over Si, would be somewhat reduced. Additionally, the efficiencies of Si systems will still rise, although they are beginning to hit practical limits,[20] and Si module costs are anticipated to decrease, but will similarly asymptote eventually as the technology is already quite mature. We were not able to make quantitative comparisons of the installed system costs in these scenarios due to the lack of PV cost projections and general uncertainty around future system cost structures, including the cost contribution of Si PV modules and the relative cost of BOS components that scale with area, which drives the impact of efficiency on installed cost. Finally, Fig. 3 includes only the costs associated with single and dual junction III-V cells; HVPE could potentially allow for the addition of even more junctions at a low cost, with the development of processes for depositing the required materials, increasing efficiencies further and enabling additional BOM and BOS cost savings out of reach for Si.

**Development of D-HVPE for high efficiency devices**

HVPE was successfully employed in the past to produce commercial devices, such as LEDs and detectors, but the production of HVPE-grown III-V devices paled in comparison to MOVPE grown devices[21,22], primarily due to the difficulty of making abrupt, highly-passivated heterointerfaces, as noted earlier. This is important because unpassivated interfaces have dangling atomic bonds and/or impurities that would otherwise act as non-radiative recombination sites and decrease device performance. It is important that the potential low costs of HVPE growth detailed in the previous section be viewed in the context of achievable performance. The key test for D-HVPE is to effectively passivate III-V layers to decrease carrier recombination at interface states, while still maintaining the high throughput that helps to make HVPE a low-cost technology.

A simple device structure that acts as a sensitive test of interface abruptness is the Esaki diode, or tunnel-junction.[23] Tunnel junctions are used in frequency converters, detectors, oscillators, amplifiers and switching circuits. This diode uses a highly-doped n-type region in intimate contact with a similar p-type region in order to enable carrier tunneling from the valence band of one side of the junction to the conduction band of the other. The change from n-type to p-type needs to occur on sub-nanometer length scales to ensure significant wave-function overlap, making the observance of tunneling behavior a good measure of interface abruptness in these epitaxial layers. Our D-HVPE reactor allows us to form GaInP/GaAs heterointerfaces both at high growth rate and without resorting to a detrimental pause in the growth process for a change in material chemistry. Growth interrupts are times when impurities can adsorb on the surface and native defects can form, both of which lead to imperfect interface passivation, increased interface roughness, and detrimental device performance such as, lack of tunneling, or a decreased solar conversion efficiency. In addition, interrupts reduce the throughput of the process, increasing device costs. Fig. 4 shows the current-density vs voltage characteristics of a tunnel junction grown via D-HVPE. This tunnel junction achieved a peak tunneling current of 11.2 A/cm$^2$, validating the ability of D-HVPE to form abrupt doping profiles. This device not only has an abrupt doping profile but also an abrupt material change from GaInP to GaAs. The extremely low resistance across it (Fig. 4 insert) allows us to use this tunnel diode in multijunction solar cell structures to connect two subcells in series such that the voltages of the individual subcells add together to create a structure with high solar conversion efficiency.[24] This indicates that D-HVPE is capable of creating heterojunctions that are thin, chemically abrupt, and free of detrimental defects that would lower the achievable performance necessary for high-peak-current tunneling.

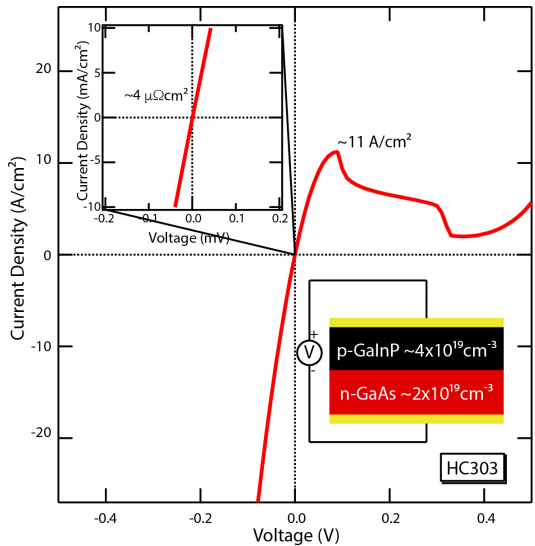

Fig. 4: Current-density-voltage characteristics for a tunnel junction grown using D-HVPE. Insert shows an amplified look around 0 V showing the low resistance of the tunnel junction.

Solar cells also require low-defect interfaces to achieve high performance and their performance provides an excellent evaluation of material quality. For this reason, we grew a series of III-V solar cells to test the ability of D-HVPE to create interfaces that reduce minority carrier recombination. Fig. 5 (left) shows a comparison of internal quantum efficiency measurements, the ratio of unreflected photons to those that are collected as useful current, of single-junction GaAs solar cells with different levels of surface and interface passivation. The details of both the growth and the processing of these devices are available in the Methods section. The addition of front and back surface passivation (from solid red to black) directly improves carrier collection at all wavelengths of light by minimizing the carrier recombination rate at these interfaces. Fig. 5 also shows for comparison an MOVPE-grown GaAs solar cell with GaInP front and rear passivation that exhibits nearly identical quantum efficiency to the D-HVPE-grown passivated device. Each of these passivated devices reaches nearly unity internal quantum efficiency over a wide range of wavelengths, from the GaAs bandedge at ~890 nm until the GaInP passivating layers begin to absorb light below ~ 670 nm, meaning that every photon absorbed in this wavelength range is converted into useful current. The nearly identical performance between the MOVPE and D-HVPE grown devices occurs despite the fact that the D-HVPE device growth rate was an order of magnitude higher than for the MOVPE control sample. This indicates D-HVPE capably creates heterointerfaces with equal performance to the incumbent batch growth technology in a continuous process that can dramatically increase III-V deposition throughput.

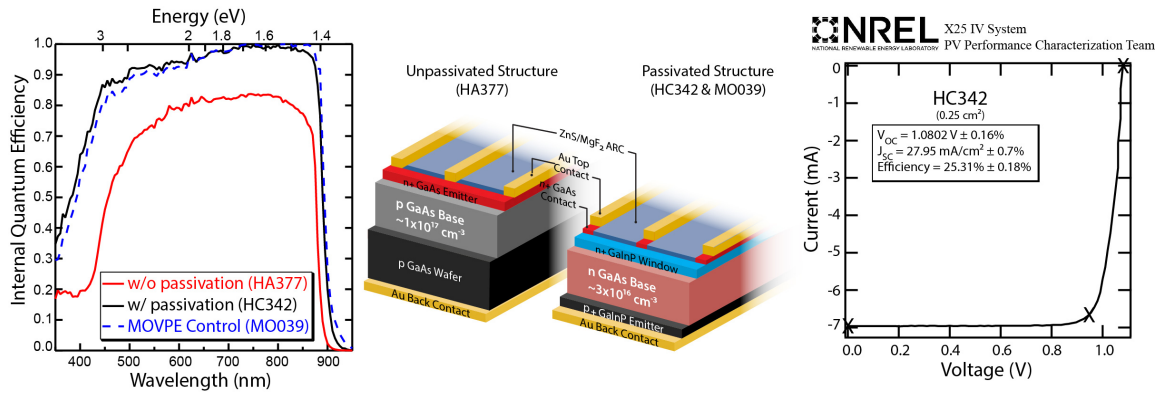

Fig. 5: (Left) Internal quantum efficiency measurements of an unpassivated (red), and a fully passivated (black) single junction GaAs solar cell grown by D-HVPE at ~ 60 μm/h. The dashed blue curve represents data from an MOVPE-grown device with nominally the same structure (shown in middle), but using a growth rate of ~ 6 μm/h. (Right) NREL-certified current-voltage data for the D-HVPE-grown GaAs solar cell with 25.3% conversion efficiency.

Fig. 5 (right) likewise shows the current-voltage characteristics of a D-HVPE-grown single-junction GaAs solar cell with front and rear passivation under simulated one-sun AM1.5G illumination. The NREL-certified measurement indicates a 25.3% conversion efficiency for a device grown at 60 μm/h. The device achieved a current density of ~28 mA/cm$^2$, which matches the current density of the control MOVPE device grown with the same structure, and is only 6% lower than the record MOVPE device which utilized a more transparent AlInP window layer[2]. The open-circuit voltage ($V_{OC}$), which can be used as a sensitive indicator of the crystal perfection in a solar cell, is 1.08 V, only 0.04 V below world-record MOVPE-grown devices.[1]

III-Vs also have the added benefit of easily incorporating additional alloy stacks to create multijunction devices that convert a larger portion of the solar spectrum even more efficiently. These devices are significantly more complicated due to the larger number of layers needed in order for the device to operate properly. We previously showed GaInP top subcells with good bulk material quality that can be coupled with the GaAs bottom subcell and the tunnel junction described here.[24] In order to demonstrate the viability of D-HVPE to manufacture more complex multijunction solar cell devices with the potential for >30% efficiency, we created a GaInP/GaAs two junction device with a GaInP/GaAs tunneling interconnect to produce a monolithic two terminal solar cell device. This device utilizes eight different layers (see Fig. 6) that need to have the right composition, doping, and low-defect interfaces; a structure that would be nearly impossible to grow by traditional HVPE. Fig. 6 shows the current density-voltage characteristics of the a multijunction solar cell device grown via D-HVPE, measured under one-sun illumination. The unpassivated top cell limits the current, however the device still increases the $V_{OC}$ from the single junction case from 1.08 V to 2.40 V, highlighting the excellent bulk quality of both subcells. This device showcases the ability of D-HVPE to grow high quality devices that utilize multiple layers of different composition and doping, while maintaining high throughput. We expect that these devices, when properly optimized, can be grown in under 5 min as opposed to the multiple hours required by MOVPE growth. We further expect that two-junction, D-HVPE-grown solar cells will yield conversion efficiencies close to 30% in the near future with simple structural modifications,[24] far in excess of the capabilities of Si PV, with a growth technique that can approach the costs of Si production.

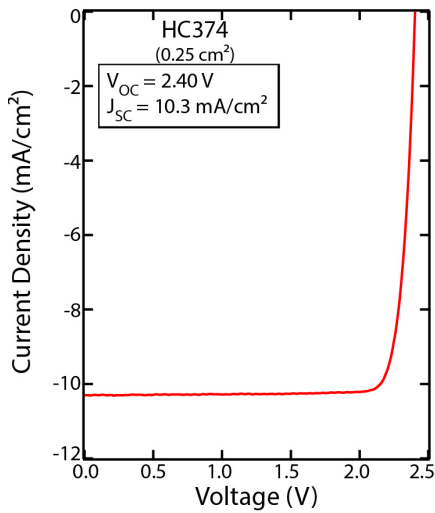
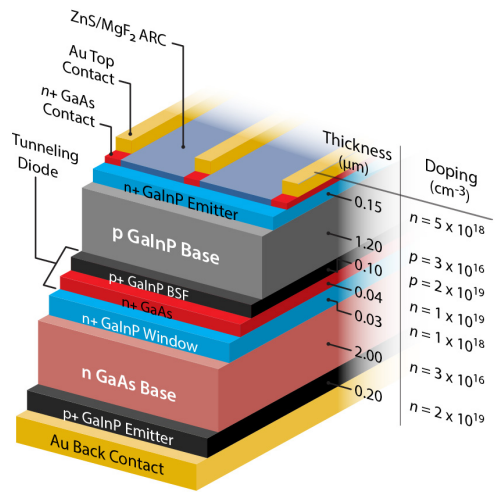

Fig. 6: Current density-voltage characteristics for an HVPE-grown multijunction solar cell device under the AM1.5G solar spectrum. The structure of device is shown on the right.

All the device results presented here show that D-HVPE is enabling for III-V device growth using this low-cost technique. We also expect the efficiency of the preliminary D-HVPE-grown solar cells shown above to increase with further optimization. For example, little work has been done to optimize the bulk material quality through the use of optimized substrate temperature or reactant flow ratios in order to decrease native or extrinsic defect concentrations. Also, the use of a more transparent AlInP window would create a significant positive increase on the current density, roughly ~ 1.2% absolute in a GaAs single junction solar cell, and >30% in a multijunction device. The addition of a AlInP window will also result in a small increase in the $V_{OC}$.. Significant fundamental research is required to determine whether high quality AlInP can be grown by HVPE, however. These improvements will lead to equal, or at least nearly-equal, device performance for a growth technique that projects to be much cheaper than existing III-V technologies.

While this section focused on recent efforts to produce III-V solar cells, the D-HVPE technique is clearly applicable to many kinds of devices. HVPE is already used to produce high-quality nitride-based template layers, as well as $Ga_2O_3$ for power electronics applications. The use of the D-HVPE process for the formation of more complicated device structures that rely on abrupt and electrically ideal interfaces will allow for the production of low-cost, high-quality transistors, light emitters and detectors, and power conditioning devices, in addition to solar cells. Indeed, the development of low cost III-V growth may enable technologies yet to be created.

**Conclusions**

We demonstrated the ability for D-HVPE to grow high efficiency devices with equivalent performance to that of conventional MOVPE, validating this new technology for the manufacture of high-performance III-V opto-electronic devices. We also developed a cost model for D-HVPE in a high volume production context. Our cost model further indicates that the use of D-HVPE together with a low-cost substrate approach could at last enable the use of III-V's in terrestrial PV markets, including certain residential and commercial rooftop installations. This technology can also be applied to manufacture non-PV devices such as high-efficiency LEDs and and devices for wireless communications applications.

*Acknowledgments*

The authors would like to thank David Guiling for materials growth and Michelle Young for device processing. This work was authored by Alliance for Sustainable Energy, LLC, the manager and operator of the National Renewable Energy Laboratory for the U.S. Department of Energy (DOE) under Contract No. DE-AC36-08GO28308. Funding provided for the cost model, tunnel junction and multijunction cell work was provided by the Advanced Research Projects Agency (ARPA-E), US Department of Energy, award #15/CJ000/07/05 and U.S. DOE Office of Energy Efficiency. Work on the single junction GaAs was provided by U.S. DOE office of Energy Efficiency and Renewable Energy, Solar Energy Technologies Office. The views expressed in the article do not necessarily represent the views of the DOE or the U.S. Government. The U.S. Government retains and the publisher, by accepting the article for publication, acknowledges that the U.S. Government retains a nonexclusive, paid-up, irrevocable, worldwide license to publish or reproduce the published form of this work, or allow others to do so, for U.S. Government purposes.

*Contributions*

*JS* helped in designing and building the D-HVPE system, processed and helped characterize the devices presented here.

*KLS* helped in the design of the D-HVPE system and in the growth and characterization of the GaInP materials.

*KH* helped develop the cost modelling.

*TR* helped develop the cost modelling.

*DLY* assisted in the design and building of the D-HVPE system.

*AJP* is the principal investigator of this work, helped in the design and building of the D-HVPE system, and contributed to the design and characterization of the tunnel junction devices.

All authors contributed to the writing in this manuscript.

**Methods**

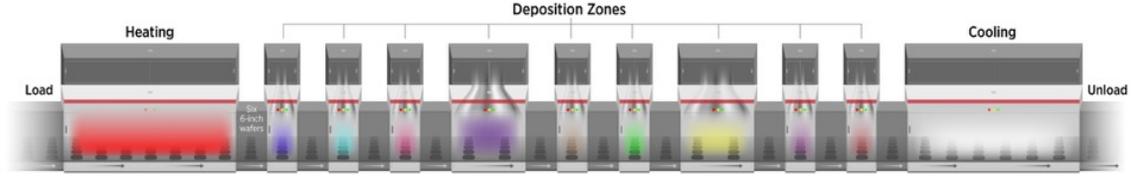

Methods Figure 1. Diagram showing the concept of the high-volume, in-line HVPE tool used for cost analysis. In our model, there is one deposition zone per layer in the device.

*D-HVPE reactor cost model*

In this model, the length of each deposition zone, *L*, is calculated as follows:

$$L_{zone\ i} = max \begin{cases} D_s \\ \frac{t_i}{R_i} \cdot 60 \cdot S_b \end{cases} \quad (1)$$

where $t_i$ is the thickness of layer *i* in μm, $R_i$ is the deposition rate for layer *i* in μm/h, $S_b$ is the reactor's belt speed in mm/min, and $D_s$ is the substrate diameter or width in mm.

We calculate the length of the heating zone similarly, except with *t* equal to the required substrate temperature (650°C) and *R* equal to the heating rate (assumed to be 130°C/min, based on heating rates currently observed in the laboratory reactor). The total length of the reactor is then equal to:

$$L_{total} = L_{heating} + L_{cooling} + \sum_{i=1}^{N} L_{zone\ i} + N \cdot L_{buffer} + 2 \cdot L_{load/unload} \quad (2)$$

where $L_{heating}$ and $L_{cooling}$ are the lengths of the heating and cooling zones, respectively, *N* is the number of deposition zones, $L_{buffer}$ is the length of the buffer zone between deposition zones, and $L_{load/unload}$ is the length required for loading or unloading. We assume $L_{buffer}$ = 0.3 m and $L_{load/unload}$ = 1.5 m. Here, we use a belt speed of 150 mm/min; this speed is within the range previously demonstrated for commercial vapor phase deposition of CdTe[25] and atmospheric pressure chemical vapor deposition (APCVD) of $Al_2O_3$[26] of 12.7 to 381 mm/min. For high volume production, we assume that there are six parallel 6" wafer tracks in the tool; similar numbers of wafer tracks have been used for commercial deposition of other thin film materials in the past. Belt speed, tool length, tool width, and deposition rate are free parameters that can be adjusted (within certain constraints) to optimize the design when a commercial reactor is developed.

We estimate that the cost of each deposition chamber would be similar to that of a continuous APCVD system[27,28], which sometimes also consist of gases flowing vertically through quartz reaction tubes onto a substrate conveyed along a belt. We collected data on single-chamber APCVD reactors via industry interview, and averaged these to obtain a base price, $P_{base}$, of $825,000, excluding automation and auxiliary costs. For a scaled HVPE reactor with similar width, we then approximate the total tool cost as:

$$C_{tool} = P_{base} \cdot \left( \frac{\sum_{i=1}^{N} L_{zone\ i}/N}{L_{base}} \right)^{\alpha} \cdot (1 + \beta \cdot (N-1)) \tag{3}$$

where $L_{base}$ is the length of the deposition area for a single chamber tool used as a proxy for HVPE costs, in this case a continuous APCVD reactor, $\alpha$ is a scaling factor for tool price with length, and $\beta$ is a scaling factor representing the fractional price of each additional zone compared to the base price. Both $\alpha$ and $\beta$ contain significant uncertainty at this point; however, based on interviews with several major suppliers of high volume deposition equipment, we approximate tool costs using $\alpha = 0.75$ and $\beta = 0.6$. Sensitivity analysis showed that the cost advantages of HVPE deposition are robust to a range of $\alpha$ and $\beta$ values. Key assumptions made in the deposition cost analysis are summarized in Table 1. Material price values are based on quotes received from material suppliers, interviews with members of industries that purchase the materials, and (where relevant) aggregation of data from online metal pricing sources and the U.S. Geological Survey. All material pricing is based on the assumption of high volume production and large orders. For HVPE deposition, we assume additional costs for automation and auxiliary equipment and installation equal to 22% and 20% of the total tool price, respectively. The material utilization rates for MOVPE are based on prior work in Ref. [10], and have been validated again by some industry members as recently as 2018. Material utilization rates for HVPE are currently uncertain. A Ga and In utilization of 70% is calculated from our research scale HVPE reactor. For all analysis, we assume U.S. manufacturing and 100% plant capacity utilization.

Table 1: Key Input Assumptions for the HVPE Deposition Cost Model

| Input | Value |
| --- | --- |
| Ga price (6N) | $0.21/g |
| In price (6N) | $0.82/g |
| High purity $AsH_3$ price | $0.48/g |
| High purity $PH_3$ price | $0.55/g |
| Ga material utilization | 70% |
| In material utilization | 70% |
| $AsH_3$ material utilization | 30% |
| $PH_3$ material utilization | 30% |
| $H_2$ curtain flow rate | 10,000 sccm |
| HCl carrier gas flow rate | 14.5 sccm |
| Tool length | 7.3m (1J cell), 9.6m (2J cell) |
| Tool price (including automation and auxiliary equipment) | $4.7million (1J), $11.6 million (2J cell) |
| Equipment maintenance cost | 4% of total equipment cost/year |

*D-HVPE III-V solar cell cost model*

We model cell costs using the NREL cost model for single and dual junction III-V solar cells under one sun illumination, first published in Ref. [10]. The cost model was refined and updated in 2018 in order to reflect current pricing for equipment and materials, as well as changes in process or capability. Fig. 5 shows the device stacks modelled here. Low-cost metallization compatible with processing and epitaxial lift-off (ELO) of III-Vs is still under development. Our cost model assumes the cell contacts are fabricated using low-cost plating without the use of gold. We also assume high volume ELO of the GaAs substrate and ≥100 substrate reuses without requiring chemical-mechanical polishing (CMP). The ability to reuse the substrate with limited or no CMP through the use of a series of buffer layers has been demonstrated in the literature, but this has not yet been demonstrated at scale or with hundreds of wafer reuses,[16,17] and this may be challenging to achieve with high yield. However, active research is taking place on substrate reuse, including chemical lift-off and mechanical fracture technologies, and a low-cost process with a high number of reuses may be feasible in the long-term.

We further assume that III-V cells can be dropped into standard modules employed for c-Si cells. We use NREL's module cost model to calculate module cost and minimum sustainable price (MSP) (see Refs. [8-11] for a description of MSP). The module MSP is input into NREL's system cost models for residential, commercial, and utility scale systems[29] to evaluate total system cost.

*D-HVPE III-V material growth methods*

All materials and devices shown in this work were grown in our dual chamber, Dynamic-HVPE (D-HVPE) reactor[30] using pure Ga and In metal, HCl, $AsH_3$, $PH_3$, and $H_2$ carrier gas. Dilute $H_2Se$ was the n-type dopant, while dimethylzinc was the p-type dopant. Heterointerfaces were formed by rapid mechanical transfer of the substrate between the two growth chambers, with each chamber stabilized at a new growth condition (either a change in material, doping, or both) for at least one minute before transfer. Substrates were (100) n+ GaAs doped with Si and offcut 4° towards the (111)B plane. The growth rates of the GaAs absorber layers was 60 μm/h, while the tunnel junction GaAs and GaInP layers were grown at 6 μm/h. The GaInP absorber in the tandem device was grown at 54 μm/h. An n+ GaAs buffer was grown to bury contamination at the initial growth interface before device growth.

Solar cell devices were processed using the method detailed in ref. [31]. Unpassivated devices were processed on wafer, while passivated devices were grown inverted and removed from the wafer. First, a broad area Au contact was electroplated onto the back-contact layer. The Au surface was bonded to a Si handle using an epoxy and the substrate was selectively etched away using a $NH_4OH/H_2O_2/H_2O$ 1:2:2 solution, exposing a GaInP etch stop layer. This layer was removed selectively with hydrochloric acid. A grid pattern was defined by standard lithography techniques, and Ni/Au front contact grids were electroplated for the front grid contacts of all devices. Finally, 5 mm x 5 mm area devices were defined by lithography and isolated using selective wet chemical etching.

We measured solar cell external quantum efficiency (EQE) on a custom instrument in which chopped, monochromatic light was split and then sent to the device of interest and a calibrated, broadband reference diode. We measured the output current of the device and reference on a lock-in amplifier, and used it to calculate the EQE, which is the ratio of electron current out to incident photons. Specular reflectance from the device surface was measured with a separate, calibrated reference diode.

We compared the measured EQEs to those of calibrated GaInP and GaAs reference cells to calculate spectral correction factors for the AM1.5G spectrum. We set the height of a Xe-lamp solar simulator to obtain an illumination of 1000 W/cm$^2$, determined by measuring the current from the reference GaAs cell held under the lamp and adjusting by the spectral correction factor for the subcell of interest. We set dual junction top cell illumination by placing the GaInP reference cell under the lamp at the GaAs one-sun height and added current using a 470 nm LED to obtain one-sun equivalent illumination. The tandem cell one-sun *J-V* curves were measured under the adjusted spectrum.